# Free and Open-Source Software is not an Emerging Property but Rather the Result of Studied Design

Paolo Magrassi, info@magrassi.net

Version 2.1 November 2010

**Abstract:** Free and open source software (FOSS) is considered by many, along with Wikipedia, the proof of an ongoing paradigm shift from hierarchically-managed and market-driven production of knowledge to heterarchical, collaborative and commons-based production styles. In such perspective, it has become common place to refer to FOSS as a manifestation of collective intelligence where deliverables and artefacts emerge by virtue of mere cooperation, with no need for supervising leadership. The paper argues that this assumption is based on limited understanding of the software development process, and may lead to wrong conclusions as to the potential of peer production. The development of a less than trivial piece of software, irrespective of whether it be FOSS or proprietary, is a complex cooperative effort requiring the participation of many (often thousands of) individuals. A subset of the participants always play the role of leading system and subsystem designers, determining architecture and functionality; the rest of the people work "underneath" them in a logical, functional sense. While new and powerful forces, including FOSS, are clearly at work in the post-industrial, networked economy, the currently ingenuous stage of research in the field of collective intelligence and networked cooperation must give way to a deeper level of consciousness, which requires an understanding of the software development process.

**Key words**: Open source – FOSS – software process – collective intelligence – Wikipedia

## 1. Introduction

One of the most intriguing aspects of the post-industrial society is the phenomenal ease and the steeply decreasing cost of human collaboration, thanks to the internet, the world-wide web and a number of tools sitting upon them, such as wikis, social networks, peer-to-peer exchanges, and mobile devices. This compounds with another mega trend, which had manifested itself earlier (Drucker 1983), i.e. the increasing importance of information and knowledge manipulation as opposed to a main focus on the transformation of physical artefacts –which was characteristic of the industrial era.

According to a growing number of authors, the combination of collaboration and dematerialization may give raise to a radically new economy typified by a combination of the following (in varying degrees and slightly different shades depending on the individual views):

- A culture of collaboration, sharing and openness will grow rapidly and in the long term possibly even replace the current atmosphere, dominated by exclusivity, closure, individualism and asymmetry (Bauwens 2005, Baldwin 2009);

- People work will be motivated less by monetary or tangible compensation and more by personal affirmation, voluntariness, ludic payoffs, and willingness to participate in attractive or important ventures (Kane 2003, von Hippel 2003, Schroer 2009);

- Enterprises need not be stable constituencies lasting years or decades, kept together by hard pacts difficult to modify: they will form opportunistically (Berkman 2009) on a project basis (Tapscott 2006), and any person may be working for any number of them at any given time;

- Enterprise organization will be less and less based on designation, co-optation and hierarchy: participation, self-determination and heterarchy prevail (Lerner 2005, Fairtlough 2005);

- A much larger number of individuals than today will actively participate with direct personal involvement in both the rule of society and the management of enterprises (Lévy 1994);

- A careful exploitation of collective intelligence may lead to a world of prosperity and peace (Lévy 1994, Tovey 2008, Baldwin 2009).

This enticing vision, in all of its variants, is certainly not without merits and does hint to issues that are at the core of the internet-enabled society, its future and the formidable opportunities it purports. However the vision is still looking for admission in the official economic circles, due to the lack of a theory showing how a cooperative, open, "sharing" economy could deliver at sustainable levels of production and growth (GDP).

Authors who have attempted to set the stage for such a theory include, but are not limited to, Eric von Hippel (von Hippel 2005), Don Tapscott (Tapscott 2006, Tapscott 1997), Yochai Benkler (Benkler 2006), Jonathan Zittrain (Zittrain 2008), Lawrence Lessig (Lessig 2008). Rather than attempting an exhausting, and hardly exhaustive, review of scientific papers published on the subject, reference to the above books, all best sellers revered in many intellectual circles, universities, research centres and symposia, will suffice to provide an account of the ongoing efforts aimed at defining the bottom-up forces that are shaping the post-industrial knowledge economy.

One thing that those publications have in common, regardless of their different flavours and approaches to describing or advocating a world of sharing and peer-production, is the assumption that free and open source software (FOSS) emerges as the product of a collective intelligence phenomenon with no need for top-down coordination and oversight. In Benkler's words, for example:

"Free software offers a glimpse at a more basic and radical challenge. It suggests that the networked environment makes possible a new modality of organizing production: radically decentralized, collaborative, and non-proprietary; based on sharing resources and outputs among widely distributed, loosely connected individuals who cooperate with each other without relying on either market signals or managerial commands." (Benkler 2006, p. 60) "Based on our usual assumptions about volunteer projects and decentralized production processes that have no managers, this was a model that could not succeed. But it did" (Benkler 2006, p. 66).

We contend that this interpretation is flawed because spontaneous participation does not remove the need for leadership in software development. We will show that large software products, whether FOSS or proprietary, are all distributed and cooperative in nature, and do require top-down controls.

## 2. Free and open-source software development: some metrics

Useful sources concerning both FOSS development metrics and the FOSS methodology and process model can be found within the FOSS community itself.

Concerning development metrics, for example, Kroah-Hartman (2008) provides us with relatively recent data. According to this source, in the 2005-2007 time frame the Linux Kernel was attended to by approximately 3,700 individuals (not all simultaneously), 86% of which were employed or contracted by enterprises and 14% were moonlighters working for free.

This statistic tells us two things. The first is that the total of Linux Kernel developers amounted on average to about 1,200 people on any given year between 2005 and 2007. It should be noted that 1,200 developers per year, some of which (as a minimum, presumably the 14% freelancers) working only part-time, is not a particularly large number by software development standards. For example, on average each of the top 25 banks in the world mobilizes a development staff of at least that size every year, developing roughly 2 million lines of code (LOC's) of software (Gartner 2009).

The second thing that the Kroah-Hartman (2008) statistic tells us is that the spontaneous, entirely autonomous participants in Linux Kernel development are a minority: six in every seven developers work as employees or contractors of ordinary businesses.

These numbers will not surprise senior software people: every experienced person who has performed software development or project management in less than trivial projects thinks that a system of 11+ million LOC's like, e.g., Linux Kernel in its 2.6.30 version (Christianson 2008, Bos 2007, Leemhuis 2009), can not be built by thousands of developers "without relying on managerial commands".

In fact, as we saw, 86% of Linux Kernel developers, the core of FOSS, work for a pay in the ranks of a company, with all the usual managerial controls. But this is not even the point. If a team of entirely independent and autonomous developers, working for free in their spare time, wanted to cooperate to the building of a software system, whether FOSS or proprietary, they would still have to submit to the requirements of design and coordination, as we will see in sections 3 and 4.

## 3 The software development process

Software is a labour intensive activity. Statistics vary greatly, but according to accurate reviews (Magrassi 1996) each "function point" of software needs between 0.5 and 2 person-days of work across the full first-cycle of a product (from conception to first user acceptance test): this implies that a software written in C, like Linux, requires in the neighbourhood of 1 person-day to get 15 working LOC's done (the number of LOC's per function point depends, among other things, on the programming language under consideration).

A large software system, such as an operating system like Linux or a full-scale business application like a bank's information system (in order of magnitude, about the size of the Linux Kernel today) amounts to millions of LOC's: therefore, in order to be delivered within a reasonable time frame, it can only be built by many hundreds, when not thousands of professionals.

### 3.1 Integration vs. modularity

This requires coordination, otherwise the system as a whole turns out incoherent. Worse yet: unlike a system such as Wikipedia, which can have low coherence but still perform decently, a software must be integrated. In Wikipedia, an article may be very good even if some of the links departing from it point to badly written or inaccurate articles. Over time, those articles will presumably be strengthened and the overall "system" (the collection of all interconnected articles) will be better: in the meantime though, it will have worked correctly in at least some of its parts. But a collection of software programs can stop working if any of the programs is bad.

One can consult a correct and informative encyclopaedia entry on "William Shakespeare" even if the related (and linked to) entries on "Stratford-upon-Avon", "Christopher Marlowe" and "Titus Andronicus" are missing or incorrect. However, one cannot run an order entry program if the related inventory- and customer-management programs are not working or inexistent. In software terminology, this is expressed by saying that Wikipedia is a loosely-coupled system, while software is tightly-coupled.

Maximizing integration (which requires coupling) and modularity (which requires mutual independence) at the same time has been the holy grail of software development since the 1960's (Böhm 1966, Dijkstra 1968, Constantine 1979, De Marco 1979). Integration favours coherence, consistency and performance, while modularity eases maintainability, increases robustness and resilience, and allows for smoother division of work.

The two goals, however, are conflicting, and only suboptimal solutions can be aimed at. This is a fact that escapes the attention of most non-software authors in the peer-production and networked economy literature, including those —such as (Baldwin 2005)– who recognize the importance of modularity in software development (although they seem to think of it as an exclusive prerogative of FOSS).

Modern software architectures, conceived specifically for highly-distributed and web-based systems, try to confront the harsh reality of software modules interdependence in various ways. Service-based architectures, for example, strive to make systems as loosely-coupled as possible, in order to limit the negative effects of missing modules and corrupted links. Direct references made by programs to one another are reduced by maintaining directories of "services" (programs). When a program needs a service it will issue a request by simply naming the service; the caller program ("consumer") needs not be linked in the same computer memory as the service's ("provider"), and the service may reside anywhere in the internet.

### 3.2 Needs for top-down supervision

This approach, however, still leaves two issues open.

To begin with, it only removes the lighter mutual-dependency problems: it does unbundle software and hardware, and it does encourage modularity; however, when a programmer sets out to write a [consumer] program, they will still need to know what provider programs do, and which parameters they must be passed, in order to make any use of their services. This requires stability of provider programs' specifications: a goal that conflicts with the fact that many programs are providers and consumers at the same time. It follows that modularity is difficult to achieve without top-down coordination, especially when the needed service belongs in a subsystem very far from the one that the programmer is working on.

Secondly, service-based architectures (like any architecture, for that matter) require stiff coordination in building directories and keeping a coherent nomenclature for all implied objects, such as programs and data sets. This is definitely a goal requiring top-down supervision. It does not matter whether such supervision is carried out by individuals or committees, since in either case people ought to be named and assigned to the coordination task: entities at a higher level than the individual programmer need to exist and exert their powers if the software is to behave coherently.

### 3.3 Feedback loops

Furthermore, while it can be relatively easy to state in plain English the general, global purpose of a software (e.g.: "A new production management system using RFID tags for components tracking and assembly", or "Adding iPhone support to Linux"), precision becomes paramount as the development project proceeds, because most computers notoriously need detailed instructions to perform even elementary tasks. The Linux Kernel, for example, is written mainly in C, a programming language much closer to hardware assembly than to human language. The C instructions for printing "hello, world" on a computer screen look as follows

```
#include <stdio.h>
int main(void)
{
    printf("hello, world\n");
    return 0;
}
```

from which the profane reader gets a grasp of how complicated it can get to instruct a computer to do such exoteric things as inventory management or shop-floor components assembly. At these levels of precision, required in the Linux Kernel like in any other software, little can be left to improvisation. It is challenging to cooperate on the creation of even a single program, amounting to a few hundreds lines of code: a comma or a bracket omitted or removed by programmer "B" may make the program obscure to programmer "A" and generate a complete misunderstanding on the side of the computer (compiler).

Implementation details are extremely important in computer programming ("coding"), and sometimes they make it impossible to comply with a given design specification, creating feedback loops that reflect backwards from subsequent to antecedent implementation stages of the project: design decisions (including the naming conventions we alluded to above) must be modified due to issues brought up at coding time and unimagined before.

This fact, referred to in software engineering by saying that the development process has the shape of a spiral (Boehm 1986), clashes with the wish of assigning to each participant developer a clear and defined task once for all and then simply waiting for his deliverable. Furthermore, and more importantly, it makes it particularly challenging to coordinate mutually-invoking programs.

### 3.4 The need for top-down design

Consider programmer John writing program P1 and programmer Mary writing P2. If they are coding, it means that some prior decision has been made that there will be a program called P1 performing certain functions, and a program called P2 performing other functions: this is called system design, or sub-system design in case P1 and P2 participate in a larger piece of software. System design decisions, in this case, might have been taken by John and Mary cooperatively, "with no managerial commands". But what about the design of a system like Linux Kernel, counting programs by the thousands and dozens of different subsystems each requiring its own design, modules split and coordination with other subsystems?

The structure of a large and complicated system can, with some simplification, be depicted as an upturned tree, from a root module (e.g., "Linux") all the way to leaves corresponding to elementary modules/programs needing no further split. $P_{jk}$, with $1 \leq j \geq M$ and $1 \leq k \geq N$, is the generic program module being attended to by programmer $A_i$, with $1 \leq i \geq R$. $R$ is the total number of programmers available and *MxN* is the dimension of the hierarchical graph representing the system structure. *M* is the breadth and *N* the depth: *j* is the number of peer modules to $P_{jk}$ (all needing to be coordinated, i.e., co-designed, along with it), while *k* measures its degree of seniority; the lower is *k*, the larger the number of modules underneath, because we are moving towards the root. It is easy to see that while a leaf module $P_{jN}$ may only need 2 people to be designed, a very high-level module may require hundreds of co-designers (all individuals who will program the modules underneath), which is obviously unrealistic.

It should by now be clear how the need for hierarchical layers of system design imposes itself: a few designers (or maybe one Linus Torvalds) at the top, then some sub-designers underneath, then sub-sub-designers, and so on. No meaningful piece of software, much less one of millions LOC's, has ever come together as a working computer program without some "higher-level" intelligence controlling the system's overall integration.

The modules attended to by individual programmers may only work together if one entity above (person, team, committee, but in any case a defined subset of the entire development team) takes care of the overall design and integration. Tales of Lego-like software componentry assembly, while attractive and suggestive, belong in the realm of software-tool vendors marketing and are inexistent in the software engineering literature. In fact, the *systems software* domain, i.e. that of Linux Kernel, is a very fortunate situation in that sense: because of the relative requirements stability, some decent degree of modularity can be achieved. But in business applications software, for example, modularization is still little more than a dream.

### 4 Organizational models in software development, FOSS or not

While explicit FOSS design supervision goes overlooked in the software-naïve literature, where coherence and design are considered as properties emerging out of a "complex system" of individuals, it is of course a very well known fact in the FOSS community.

| Al Viro | 1.9% |
| David S. Miller | 1.8% |
| Adrian Bunk | 1.7% |

| Ralf Baechle | 1.6% |
|---|---|
| Andrew Morton | 1.5% |
| Andy Kleen | 1,2% |
| Takashi Iwai | 1.2% |
| Tejun Heo | 1.1% |
| Russel King | 1.1% |
| Steven Hemminger | 1.1% |

Table 1: The top ten Linux Kernel developers in 2005-2007. Source (Kroah-Hartman 2008)

Kroah-Hartman (2008) again provides us with some information: relatively few individuals determine and even produce directly a substantial amount of the work. In 2005-2007, for example, the ten persons listed in Table 1 produced 14% of everything that was done in Linux Kernel. The top 30 individuals produced 30%. These are the people who, along with Torvalds, made the top design decisions. (The vast majority of the other 3700 people or so mainly –although not exclusively– did bug fixing: an activity, according to Raymonds (1999), at which "crowds" excel).

This does not mean that design proposals cannot be made by anyone else; it does not mean that Torvalds et al. exert managerial controls such as assigning tasks to specific people; it does not mean that participants cannot often pick their preferred tasks from a to-do list; it does not negate that many design decisions are made by committee rather than by an individual alone (exactly as it happens with proprietary software products): but it does say that most Linux (or any other FOSS product) programmers submit to design decisions made by someone "above".

In proprietary contexts, the people above often happen to be higher-ranked in a company hierarchy. But conceptually (and apart from the fact that, as we saw, the majority of FOSS developers do work in the ranks of enterprises) this difference is irrelevant to the present discussion: whatever their hierarchical positions, software designers do make decisions that the rest of the people must conform to.

Eric S. Raymond, one of the most authoritative FOSS gurus (his 1999 book "The Cathedral and the Bazaar" had received 4,370 citations in Google Scholar by April, 2010, although perhaps it is not as thoroughly read as it is readily quoted), describes the FOSS design issue in Raymonds (2000a):

"The trivial case is that in which the project has a single owner/maintainer. […] The simplest non-trivial case is when a project has multiple co-maintainers working under a single "benevolent dictator" who owns the project. Custom favours this mode for group projects; it has been shown to work on projects as large as the Linux kernel or Emacs. […]

As benevolent-dictator projects add more participants, they tend to develop two tiers of contributors; ordinary contributors and co-developers. A typical path to becoming a co-developer is taking responsibility for a major subsystem of the project. Another is to take the role of "lord high fixer", characterizing and fixing many bugs. […] A co-developer who accepts maintenance responsibility for a given subsystem generally gets to control both the implementation of that subsystem and its interfaces with the rest of the project, subject only to correction by the project leader (acting as architect). […]

By custom, the "dictator" or project leader in a project with co-developers is expected to consult with those co-developers on key decisions. [...]

Some very large projects discard the `benevolent dictator" model entirely. One way to do this is turn the co-developers into a voting committee (as with Apache). Another is rotating dictatorship, in which control is occasionally passed from one member to another within a circle of senior co-developers; the Perl developers organize themselves this way. Such complicated arrangements are widely considered unstable and difficult."

[Copyright © Eric Steven Raymond 1998]

The picture is clear. FOSS development is based on one of two organizational models: the benevolent dictator or the design committee, and the former model creates less problems. This is hardly any different from proprietary-software development.

## 5. Conclusions

The development of a less than trivial piece of software, irrespective of whether it be FOSS or proprietary, is a complex cooperative effort requiring the participation of up to thousands of individuals. A subset of the participants must play the role of system and subsystem designers, determining the systems' architecture and functionality, and the rest of the people work "underneath" them in a logical, functional sense.

This submission needs not also be hierarchical: the same applies frequently to proprietary-software development as well, where ample use is made of contractors and outsourcers of various sorts, all without hierarchical connections to system designers/architects.

All large software development projects, whether FOSS or not, are attended to by geographically distributed teams, as is typical of the very intricate ramifications of "global" business organizations today. Business software applications, for example, are typically developed (by vendors and/or user enterprises and/or professional services organizations) by teams working for dozens of different legal entities and dispersed across two or more continents. It is not uncommon to find, in such teams, a minority of free-lance, self-employed professionals. All the people involved, in any case, are accustomed to having two lines if reporting, quite often distinct: one thing is the hierarchical manager (the boss), another thing is the project supervisor or subsystem architect.

Bottom-up participation isn't FOSS-exclusive, either: in all software development contexts, individual developers can make design or organizational proposals that extend beyond the scope of their formal assignment, and this is precisely the way most people progress from programmer to higher-level roles. This escalation sometimes entails the climbing of the enterprise hierarchy as well, but even this is not obvious: many corporations allow for "professional" career paths with little hierarchical implications but still rewarding for the individual.

### 5.1 What is typical of FOSS?

And this is the point. What really characterizes FOSS, from a human resources viewpoint, is that a significant, although smaller than is usually believed (the 14% of the Kroah-Hartman statistic, in the case of Linux), num-ber of people participate spontaneously and without pay. Designers set guidelines and maintain to-do lists: some developers spontaneously pluck from those.

The literature on the motivations for participating to FOSS projects is vast. From, for example, Kane (2003), von Hippel (2003), Raymonds (2000b), and Schroer (2009) we learn that FOSS developers are motivated by the willingness to participate in an attractive and important challenge, by the fact that their names will appear on the list of software and projects owners, by a genuine sense of sharing and participation, by *homo ludens* payoffs, and sometimes by the urge to contrast dominant software players and "monopolies" like Microsoft (a urge carefully cultivated and fomented by Microsoft's adversaries, including IBM, Red Hat, Intel, Novell, Sun/Oracle, Hp and many others, who are the employers of that 86% developers working on Linux as well as of those working on all other FOSS products, and are the "hidden" market forces behind such products).

The second fact that separates FOSS from proprietary software is, of course, the novel ownership models reflected by the original General Public Licence and the many others that have been developed from it since its inception in 1989.

## 5.2 What is not in FOSS, and in proprietary software either

The notion of a coherent and performing system emerging from a crowd of spontaneous contributors without top-down direction and supervision is unfounded: it does not correspond to the way software of any kind is designed and built.

## 6 Further research

Spontaneous/voluntary participation of contributors and the new intellectual property scheme, i.e. the quintessence of FOSS, are formidable drivers of change and innovation in the post-industrial economy, with consequences and implications extending far beyond the reach of software. Economists, sociologists, jurists, political scientists, psychologists are finding and will increasingly find in "open content" a myriad of research motivations.

In the case of software, the naïve notion of systemic emergence should be abandoned and, to investigate further what makes the FOSS production model different, research should be carried out on, among other things, the following:

- What are the relationships and the interplay between hierarchical and functional dependencies in software development organizations?

- Is it easier in FOSS (than in proprietary-software development) to achieve [sub]system designer status irrespective of one's hierarchical position?

- Do bottom-up design proposals occur more frequently in FOSS than in proprietary-software development?

- Are FOSS products more modular than proprietary products, when both are considered at the same level of abstraction with respect to hardware architecture?

Finally, concerning an issue which this paper only touched upon quickly but is strictly connected to the division of labour discussion, there is a need to study what, if any, are the market drivers behind FOSS products. Most of the literature we have referred to seems to as-sume, and often explicitly states, that FOSS products are built by "loosely connected individuals who cooperate with each other without relying on […] market signals" (Benkler 2006, page 60). The role played in FOSS by the many ordinary businesses (sometimes huge software or hardware vendors) who directly or indirectly employ most of the developers must be understood better. This will provide a sharper insight into the dynamics of peer production, the culture of sharing, and collective intelligence.